\documentclass[12pt]{article}
  
\usepackage{axodraw}
\usepackage{epsfig}
\usepackage{calrsfs,bbm,bm}
\usepackage{color,cancel,graphicx,mathrsfs}
\usepackage{dsfont}
\usepackage{amsmath}
\usepackage{amssymb}
\usepackage{exscale}
\usepackage{hyperref}

\input pix.sty

\def\a0size{6}

\newcommand{\lsi}{\raise0.3ex\hbox{$<$\kern-0.75em\raise-1.1ex\hbox{$\sim$}}}
\newcommand{\gsi}{\raise0.3ex\hbox{$>$\kern-0.75em\raise-1.1ex\hbox{$\sim$}}}
\newcommand{\lsim}{\mathop{\lsi}}
\newcommand{\gsim}{\mathop{\gsi}}
\renewcommand{\vec}[1]{{\bf #1}}

\newcommand{\be}{\begin{equation}}
\newcommand{\ee}{\end{equation}}
\newcommand{\baed}{\begin{aligned}}
\newcommand{\eaed}{\end{aligned}}

\allowdisplaybreaks[1]
\newcommand{\Yuk}{int}
\newcommand{\rmi}[1]{{\mbox{\scriptsize #1}}}
\newcommand{\rmii}[1]{{\mbox{\tiny\rm{#1}}}}

\newcommand{\ba}{\begin{eqnarray}}
\newcommand{\ea}{\end{eqnarray}}
\newcommand{\eq}{}
\newcommand{\eqs}{}
\newcommand{\nr}[1]{(\ref{#1})}
\newcommand{\se}{sec.~}
\newcommand{\ses}{secs.~}
\newcommand{\nn}{\nonumber \\}
\newcommand{\Ref}{ref.~}
\newcommand{\flav}{flavour\ }
\newcommand{\col}{colour\ }
\newcommand{\gaussian}{gaussian\ }

\newcommand{\hnu}{h }

\def\procA{\pic{%
 \Lqu(0,0)(14,14)%
 \Lsc(0,30)(14,16)%
 \Line(14,14)(30,14)%
 \Line(14,16)(30,16)%
}}
\def\procB{\pic{%
 \Lqu(0,0)(9,9)%
 \Line(0,0)(14,14)%
 \Lgl(7,7)(7,23)%
 \Lsc(0,30)(14,16)%
 \Line(14,14)(30,14)%
 \Line(14,16)(30,16)%
}}
\def\procC{\pic{%
 \Lqu(0,0)(14,14)%
 \Lsc(0,30)(14,16)%
 \Agl(7,23)(5,-45,135)%
 \Line(14,14)(30,14)%
 \Line(14,16)(30,16)%
}}
\def\procD{\pic{%
 \Line(0,0)(14,14)%
 \Lqu(1,1)(14,14)%
 \Lsc(0,30)(14,16)%
 \Agl(7,7)(5,-135,45)%
 \Line(14,14)(30,14)%
 \Line(14,16)(30,16)%
}}
\def\procE{\pic{%
 \Lqu(0,0)(14,14)%
 \Lsc(0,30)(14,16)%
 \Lgl(0,13)(9,21)%
 \Line(14,14)(30,14)%
 \Line(14,16)(30,16)%
}}
\def\procF{\pic{%
 \Lqu(0,0)(9,9)%
 \Line(0,0)(14,14)%
 \Lgl(0,18)(9,9)%
 \Lsc(0,30)(14,16)%
 \Line(14,14)(30,14)%
 \Line(14,16)(30,16)%
}}
\def\procG{\picb{%
 \Lqu(0,0)(15,15)%
 \Lgl(0,30)(15,15)%
 \Lqu(15,15)(29,15)%
 \Lsc(31,14)(45,0)%
 \Line(29,15)(44,30)%
 \Line(31,14)(45,28)%
}}
\def\procH{\pic{%
 \Lqu(0,30)(15,26)%
 \Lgl(0,0)(15,4)%
 \Lsc(15,4)(15,24)%
 \Lsc(15,4)(30,0)%
 \Line(15,24)(30,28)%
 \Line(14,26)(30,30)%
}}
\def\procI{\pic{%
 \Lqu(0,30)(15,26)%
 \Lsc(0,0)(15,4)%
 \Lqu(15,26)(15,6)%
 \Lgl(15,26)(30,30)%
 \Line(15,6)(30,2)%
 \Line(14,4)(30,0)%
}}
\def\procJ{\pic{%
 \Lqu(0,30)(15,26)%
 \Lsc(0,0)(15,4)%
 \Lsc(15,4)(15,24)%
 \Lgl(15,4)(30,0)%
 \Line(15,24)(30,28)%
 \Line(14,26)(30,30)%
}}

\makeatletter
\renewcommand\section{\@startsection {section}{1}{\z@}%
                                   {-5.5ex \@plus -1ex \@minus -.2ex}
                                   {2.3ex \@plus.2ex}%
                                   {\normalfont\large\bfseries}}
\renewcommand\subsection{\@startsection{subsection}{2}{\z@}%
                                     {-3.25ex\@plus -1ex \@minus -.2ex}%
                                     {1.5ex \@plus .2ex}%
                                     {\normalfont\normalsize\bfseries}}
\renewcommand\thesection {\@arabic\c@section}
\renewcommand\thesubsection   {\thesection.\@arabic\c@subsection}
\renewcommand{\@seccntformat}[1]{%
\csname the#1\endcsname.\hspace{1.0em}}
\makeatother

\begin{document} 

\setlength{\baselineskip}{0.6cm}
\newcommand{\figysize}{16.0cm}
\newcommand{\figtopspace}{\vspace*{-1.5cm}}
\newcommand{\figbottomspace}{\vspace*{-5.0cm}}
  

\begin{flushright}
BI-TP 2014/03  
\\
April 2014 
\\
\end{flushright}
\begin{centering}

\vfill

{\Large\bfseries
 Kubo relations and radiative corrections   \\[2.5mm]
 for lepton number washout 
}

\vspace*{.6cm}

Dietrich~B\"odeker$^\rmi{a,}$\footnote{bodeker@physik.uni-bielefeld.de} and 
M.~Laine$^\rmi{b,}$\footnote{laine@itp.unibe.ch}

\vspace*{.6cm} 

$^\rmi{a}$%
{\em 
Fakult\"at f\"ur Physik, Universit\"at Bielefeld, 33501 Bielefeld, Germany
}

\vspace*{0.3cm}

$^\rmi{b}$%
{\em
Institute for Theoretical Physics, 
Albert Einstein Center, University of Bern, \\ 
Sidlerstrasse 5, 3012 Bern, Switzerland
}

\vspace*{1cm}
 
{\bf Abstract}

\end{centering}
 
\vspace{0.5cm}
\noindent
The rates for lepton number washout in extensions of the Standard
Model containing right-handed neutrinos are key ingredients in
scenarios for baryogenesis through leptogenesis.  We relate these
rates to real-time correlation functions at finite temperature,
without making use of any particle approximations.  The relations are
valid to quadratic order in neutrino Yukawa couplings and to all orders 
in Standard Model couplings. They take into account all spectator processes, 
and apply both in the symmetric and in the Higgs phase of the electroweak
theory. We use the relations to compute washout rates at next-to-leading order
in $ g $, where $ g $ denotes a Standard Model gauge or Yukawa coupling, 
both in the non-relativistic and in the relativistic regime. Even in the
non-relativistic regime the parametrically dominant radiative corrections are
only suppressed by a single power of $ g $.  In the non-relativistic regime
radiative corrections increase the washout rate by a few percent at high
temperatures, but they are of order unity around the weak scale and in the
relativistic regime.

\vspace{0.5cm}\noindent

 
\vspace{0.3cm}\noindent
 
\vfill \vfill
\noindent
 

 
%
\section{Introduction and motivation}
\label{sec:intro} 

With the Standard Model (SM) of electroweak interactions being 
validated up to an energy scale of several hundred GeV by the 
LHC experiments, and the knowledge that the SM
combined with the conventional Big Bang scenario can explain neither the 
baryon asymmetry of the Universe, nor the existence of dark matter, 
it is appealing to search for an answer to  these cosmological
questions from other than electroweak interactions. The physics
associated with right-handed (RH) neutrinos, which can also 
account for the observed left-handed neutrino mass differences and
mixing angles, is rich enough to conceivably solve both 
problems~\cite{fukugita,numsm}. In this paper we concentrate on the 
baryon asymmetry aspect, and refer to the 
scenario as leptogenesis.  

The original description of leptogenesis 
uses Boltzmann equations for the phase space density
of the participating particles as a starting point. 
These are then integrated over momenta assuming kinetic 
equilibrium. The processes considered are the decays ($1\to 2$) 
and inverse decays ($2\to 1$) of the heavy RH neutrinos. 
In addition, so-called spectator processes, 
which proceed much faster than the expansion of the Universe, 
determine which quantities are in thermal equilibrium. 
Frequently also additional scattering processes ($2\leftrightarrow 2$)
and inverse decays have been included. 

The $2\leftrightarrow 2$ processes are a part of the radiative corrections.
To assess the theoretical uncertainty of the analysis
it would be desirable to compute the
complete next-to-leading order (NLO) radiative corrections. 
It is not clear how to do this consistently
within a description based on Boltzmann equations. This has motivated 
several authors to search for a first principles description of 
leptogenesis, without already putting in the set of assumptions 
and approximations which are implicit to the Boltzmann equations. 
Among the strategies followed are Kadanoff-Baym and similar equations 
for Green's functions 
(cf.\ refs.~\cite{beneke-density,anisimov-quantum,mg}
for recent work and references). 
Although the starting point is exact, it may be 
difficult to perform systematic
NLO calculations in these settings. 

In this paper we propose a different route towards a first principles
understanding of leptogenesis. 
We first formulate a rather general
non-equilibrium problem. We argue that it can be described by a simple set 
of ordinary differential equations. 
The coefficients in these effective equations
are shown to be related to real-time correlation functions
at finite temperature. We then focus on one of the coefficients, 
the dissipation of
lepton minus baryon number $ n _ { L -B } $,\footnote{Or closely related
quantities, depending on the temperature under consideration.} 
which in the absence of expansion is described by 
\begin{align} 
  \frac{ {\rm d} n^{ } _ {L-B} } { {\rm d} t } = 
  - \gamma^{ }_{ L-B }  \, n^{ } _ { L-B }
  + O(n_{L-B}^2)
 \;.
\end{align} 
We calculate the dissipation coefficient\footnote{%
Or, again depending on the temperature, the dissipation matrix.}
$\gamma^{ }_{ L-B }$
up to NLO in the SM couplings. 
It is shown, in particular, that the dominant NLO corrections are 
only suppressed by a single power of the gauge or top Yukawa couplings,  
and have a substantial relative influence even 
at temperatures much below the mass
of the lightest right-handed neutrino. 

In sec.~\ref{sec:physical} we describe the physical picture behind
baryogenesis through leptoge\-nesis. Sec.~\ref{s:charges} contains a
general analysis of the dissipation or washout rates of almost
conserved charges, and relations of these rates to real-time
correlation functions at finite temperature. We show that these rates
factorize into a real-time spectral function and the inverse of a
susceptibility matrix. In sec.~\ref{sec:washout} we specialize to the
washout of lepton minus baryon number in extensions of the SM with
right-handed neutrinos. We obtain a master formula, and leading-order (LO)
and NLO results for its two ingredients, the spectral function and the
susceptibility matrix. In sec.~\ref{sec:asymmetry} we study the effect
of the radiative corrections to the washout rate on the lepton
asymmetry for one particular set of parameters. We summarize and
conclude in sec.~\ref{sec:summary}.

{\bf Notation} 
4-vectors are denoted by lower-case italics and  
3-vectors by boldface, and the metric is such that 
$p^2 = (p^0)^2 - \vec{p}^2$. For 
spatial integrals we use the notation 
$ \int _ { \vec x  } \equiv \int { \rm d}^ 3 x $,
space-time integrals are denoted by
$ \int _ x \equiv \int { \rm d} ^ 4 x $. 
Spatial momentum integrals are written as 
$ \int _ { \vec k } \equiv \int { \rm d} ^ 3 k /(
2 \pi ) ^ 3 $.

%
\section{Physical picture} \label{sec:physical} 

We start by outlining the physical picture for how our computation
fits in a generic leptogenesis framework
(a recent example can be found in ref.~\cite{nrlg}). 
A key observation is that in leptogenesis 
the system is almost in thermal equilibrium.\footnote{With the exception 
of leptogenesis during reheating, which is a process far from equilibrium.} 
Most physical
quantities rapidly fluctuate thermally around their equilibrium values.
The corresponding reactions are often referred to as 
``spectator processes''~\cite{buchmuller-spectator,roulet-spectator}. 
Other quantities relax to 
equilibrium on time scales much larger than the Hubble time, so 
that they can be considered conserved. A few  have 
relaxation times of the order
of the Hubble time. Only these have to be taken into account as
dynamical degrees of freedom, and we will refer to them as ``slow''. 
What they are depends on the Hubble rate and thus on 
the temperature. A non-equilibrium state is then 
characterized by deviations  of
the slowly relaxing quantities from their equilibrium values which are much
larger than a typical thermal fluctuation. 

One of the slowly relaxing quantities is $ L - B $. It is violated by 
the Yukawa interactions between Standard Model leptons and right-handed 
(gauge singlet or ``sterile'') neutrinos 
$ N _ {I } $, with Majorana masses $ M _ I $.
Therefore, if at some time in the evolution 
of the Universe $ L-B $ is non-zero, these interactions tend 
to reduce it (examples of processes are shown in \fig\ref{fig:processes}). 
There can also be a source term 
for the lepton number if the number density of 
right-handed neutrinos deviates from thermal equilibrium and 
if their interactions violate CP. This would  lead to baryogenesis 
through leptogenesis~\cite{fukugita}.

%
\begin{figure}[t]
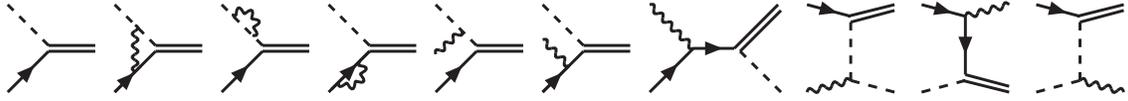


\begin{eqnarray*}
&& 
 \hspace*{-1cm}
 \procA
 \hspace*{0.45cm}
 \procB
 \hspace*{0.45cm}
 \procC
 \hspace*{0.45cm}
 \procD
 \hspace*{0.45cm}
 \procE
 \hspace*{0.45cm}
 \procF
 \hspace*{0.45cm}
 \procG
 \hspace*{0.75cm}
 \procH
 \hspace*{0.55cm}
 \procI
 \hspace*{0.55cm}
 \procJ
\end{eqnarray*}

\caption[a]{\small 
 Examples of processes, up to $O(g^2)$, through which a net
 lepton number can be ``washed out''. 
 Arrowed, dashed, and wiggly lines correspond to 
 leptons, scalars, and gauge fields, whereas
 RH neutrinos are denoted by a double line. Additional reactions
 (not shown) involve the generation of antileptons or processes mediated
 by SM Yukawa or Higgs couplings.  
} 
\la{fig:processes}
\end{figure}
%

If the slow variables have only small deviations from equilibrium, 
we can linearize their equations of motion. 
For example, assuming that the only slow degrees 
of freedom 
are the lepton minus baryon number 
density $n^{ } _ {L-B} $ 
and the right-handed neutrino phase space distribution 
$f^{ } _ {\! N}
$ (momentum, spin and 
\flav indices are suppressed and it is assumed 
that the entanglement of the right-handed neutrinos does not play a role),
the evolution equations are of the form
\ba
 \mathcal{D}_t\, f^{ }_{\! N}
 & = & - \gamma_N^{(2)}
         \, \bigl[ f^{ }_{\! N}
         - f_\rmi{eq}  \bigr]
       - \gamma_ { N, L-B }^{(4)} \, n^{ }_ { L-B } 
       + \ldots \;, 
       \nonumber 
 \\
  \mathcal{D}_t\, n^{ }_{L-B} & = & 
 - \gamma_{L-B}^{(2)} \, n^{ }_{L-B} -
   \gamma_{L-B,N}^{(4)}  
         \, \bigl[ f^{ }_{\! N} - f_\rmi{eq}  \bigr] 
 + \ldots \;, \label{rate2}
\ea
where $\mathcal{D}_t$ is the appropriate time derivative 
in an expanding background, 
$f_\rmi{eq}$ is the equilibrium distribution, 
and terms of higher 
orders in deviations from equilibrium and in time derivatives have been 
neglected.\footnote{%
 In situations where 
 the deviation from thermal equilibrium is sizeable, e.g.\ in the 
 weak washout regime, non-linear
 terms could also play a role. One would expect the dominant non-linear
 contribution to $\mathcal{D}_t\, n^{ }_{L-B}$ to be of order
 $h^2 [f^{ }_{\! N} - f_\rmi{eq}] \, n^{ }_{L-B}$. 
 }  
Within this effective description, the coefficients $\gamma$
are independent of the values of the slow variables
$n_ { L - B } 
$ and $f^{ }_{\! N}- f_\rmi{eq}$ and 
of the Hubble rate appearing in $\mathcal{D}_t$. In particular, 
they can be determined arbitrarily close to equilibrium and
with vanishing Hubble rate, to any order in Standard Model couplings.  
This is the philosophy that underlies Kubo relations~\cite{kubo}.

The coefficients $\gamma^{(2)}$ in eq.~\nr{rate2} 
are of $O(\hnu^2)$, 
where $\hnu $ denotes a generic neutrino Yukawa coupling. 
The terms containing
$\gamma^{(4)}$ violate CP, and must contain  additional Yukawa couplings with
makes them $ O ( \hnu ^ 4 ) $.
If the terms omitted from \eq
\nr{rate2} 
are suppressed by increasingly high powers of $\hnu $, 
and $\hnu$ is very small, we can make use of a perturbative
expansion in $\hnu $. At leading order in $ \hnu ^ 2 $, 
$\gamma^{(2)}_{N}$ can be determined by setting
$f^{ }_{\! N} = 0$ 
because the 
contributions from the omitted higher-order terms 
are of $O(\hnu ^4 f_\rmi{eq}^2)$.
Then $\gamma^{(2)}_N f_\rmi{eq}$ agrees with 
the right-handed neutrino production rate of which a lot
is known: it has been computed
up to NLO in SM couplings
in the non-relativistic~\cite{salvio,nonrel,biondini} 
and relativistic~\cite{master,relat} regimes, 
and to LO in the ultrarelativistic 
regime~\cite{anisimov,besak}, which necessitates a resummation
of the loop expansion. With a similar philosophy, 
the CP-violating source term $\gamma^{(4)}_{L-B,N} f_\rmi{eq}$
has been expressed in terms of a Green's function
which could in principle be evaluated at NLO~\cite{gagnon}. 

In the present paper, we are concerned with the coefficient 
$\gamma^{(2)}_{L-B}$, which may be called the lepton minus baryon 
number washout or dissipation rate. 
Within the range of validity of 
\eq
\nr{rate2}, it can be computed in a system 
with $f^{ }_{\! N} = f^{ }_\rmi{eq}$, $\mathcal{D}_t = \partial_t$, 
and assuming $n^{ }_{L-B}$ to be close to its equilibrium value. 
If $f^{ }_{\! N} \neq f^{ }_\rmi{eq}$, the rate 
$\mathcal{D}_t\, n^{ }_{L-B}$ changes, but the change originates
from the other terms in \eq\nr{rate2} rather than from 
a change of $\gamma^{(2)}_{L-B}$.

%
\section{General analysis}
\label{s:charges}

%
\subsection{Conserved and approximately conserved charges}
\label{ss:charges} 

As a first step one has to identify the 
spectator processes.
These processes conserve some charges, which by Noether's theorem
are associated with the symmetries that the corresponding 
interactions respect. 
Some of these symmetries will be broken
by the slow interactions, and the corresponding charges will no longer
be conserved. If some symmetries remain unbroken, there are
linearly independent charges
$  X _ {  \bar{ a } }  $ which are still conserved.\footnote{%
 We choose this set to be complete in the sense that there are 
 no additional charges which are linearly independent of 
 the $ X _ {  \bar{ a } }  $ and which are conserved. In the actual 
 analysis, it may be possible to choose the
 $ X _ { \bar a  } $  such that not all of them need to
 be included  explicitly, cf.\ the discussion below 
 eq.~(\ref{washout11}).} 
In addition, there  are charges $  X _ a $ which together with 
the $ X _ { \bar{ a } } $ form a linearly independent set,
such that the original charges can be written as linear combinations 
of the $ X _ a $ and $ X _ { \bar{ a } } $. 
By definition the $ X _ a $
are no longer conserved. Their values 
can now depend on time and one finds equations of motion for them 
of the type in \eq
\nr{rate2}. 
The choice of the non-conserved charges $ X _ a $ is not unique,
because after adding a conserved charge they are still
non-conserved. However, we consider a state in which the strictly
conserved charges vanish.
Then this ambiguity is irrelevant. 

For computing the washout rates the expansion of the Universe
can be ignored, as well as the interactions which are much slower than 
the expansion.  We assume that the thermal expectation 
values of the $ X _ a $ vanish,
$ \langle X _ a \rangle _ {\rm eq } =0 $. 
Now we consider a non-equilibrium system in which the $ X _ a $ 
start with some non-zero values, which we assume to be much larger than
their typical thermal fluctuation. The non-equilibrium state
is completely specified by the values of $ X _ a $. Thus,
their time derivatives only 
depend on their values and on the temperature. 
For sufficiently small values one can expand $ {\rm d} X _ a / {\rm d} t $
in powers of $ X _ a $, and keep only the linear term, 
\begin{align} 
   \frac{ {\rm d} X _ a } { {\rm d} t } = - \gamma  _ { a b } X _ b 
   \;. \label{eom}
\end{align} 
It turns out that 
the coefficients $\gamma _ {a b}$ 
can be factorized into
two parts (cf.\ eq.~\nr{washout3.9.0a}): to a real-time 
``spectral function'' 
$\rho_{ a c }(\omega)$,  and 
to the inverse of a static ``susceptibility matrix'', 
denoted by $\Xi^{-1} _{c b}$. 
The latter contains similar information 
as the \flav matrix $A$ introduced
in ref.~\cite{matA}, or the coefficients $c_\ell, c_\varphi$ that 
appear widely in leptogenesis literature (cf.\ ref.~\cite{davidson-review}). 
We define these two parts in turn. 

%
\subsection{Kubo relations for washout rates}

To determine the coefficients $ \gamma  _ { a b } $ 
we proceed similarly to the general 
method\footnote{For a recent application in
relativistic field theory see \Ref\cite{chemical}.} 
described in ref.~\cite{landau5}. 
We work to leading order in the neutrino 
Yukawa interaction (see below), but, if not stated otherwise, 
to all orders in other interactions. 

Even in thermal
equilibrium the values of the $ X _ a $ are not 
constant in time, but instead they 
fluctuate around their equilibrium values which we assume to be zero. 
For {\em small frequencies} these fluctuations can be described by 
an effective classical theory 
with an  equation of motion similar to \eq\nr{eom}. 
The only difference compared with \eq(\ref{eom}) 
is that on the right-hand side there is an additional 
\gaussian noise term.
This equation of motion can be solved to
write the fluctuation of $ X _ a $ at time $ t $ 
in terms of its value at time zero
plus the contribution from the noise. This can be 
used to compute the real-time correlation function  
\begin{align} 
   { \cal C } _  { a b } ( t ) \equiv 
   \langle X _ a ( t ) X _ b ( 0 )  \rangle 
   \; \label{correlation}
   ,
\end{align} 
where $ \langle \cdots  \rangle $ denotes the thermal average
over an ensemble in which 
the strictly conserved charges are zero and do not fluctuate. The 
effective description is classical, so the ordering inside the average 
does not play a role. Since the noise is \gaussian it  drops out when 
taking the noise average, and one obtains
\begin{align}
   { \cal C } _ { a b } ( t ) = \left ( e ^{ - \gamma  t } \right ) _ { a c } 
   \langle X _ c X _ b \rangle 
   \;, \qquad    t > 0  
   \;.
\end{align} 
Taking the one-sided Fourier (or Laplace) transform we obtain 
\begin{align} 
   { \cal C } ^ + _ { a b } ( \omega  )
   &\equiv  \int _ 0 ^ {\infty  } \! {\rm d} t\, e ^ { i  \omega   t } 
   \, { \cal C } _ { a b } ( t )
   \nonumber\\
   & = {}- \left ( i \omega  - \gamma  \right ) ^{ -1 } _ { a c } 
   \langle X _ c X _ b \rangle 
   \;, \label{C_clas}
\end{align} 
where   $ \omega  $ has a positive imaginary part. 

The effective description 
of the correlation function in \eq(\ref{correlation}) is
valid when $ \omega  \ll \omega  _ \rmii{UV} $, where $ \omega  _ \rmii{UV} $
is a characteristic frequency of the fluctuations of other, ``faster''
relaxation processes, the spectator processes. 
Now let $ \omega  $ be real and $ \omega  \gg \gamma  $, where $\gamma$
denotes the absolute value of the largest eigenvalue of $\gamma_ { a b }$.
Then  one can expand \eq\nr{C_clas}
and finds
\begin{align} 
   { \rm Re } \, { \cal C } ^ + _ { a b } ( \omega +i 0 ^ + )
  = \frac{ 1 } { \omega  ^ 2 } \, 
   \gamma  _ { a c } \langle X _ c X _ b \rangle 
   + O \left ( \omega  ^{ -4} \right ) 
   \;. \label{ceff}
\end{align} 

On the other hand, the correlation function in \eq\nr{correlation} 
can also be computed in the microscopic 
quantum field theory. By matching the results in the regime
$ \gamma  \ll \omega  \ll \omega  _ \rmii{UV} $ 
one can then obtain the coefficients $ \gamma  _ { a b } $, 
together with a consistency check on the functional form of 
the $\omega$-dependence. 
In quantum field theory we define the
correlation function as
\begin{align} 
   C  _ { a b } ( t  ) \equiv
   \Bigl\langle\, \frac12 \{X _ a ( t ) , X _ b ( 0 ) \}\, \Bigr\rangle 
  \;, 
\end{align} 
where the angular brackets indicate an average over a thermal ensemble 
in which the values of all strictly conserved charges are zero. Making
use of text-book relations between frequency-space
correlators involving anticommutators
and commutators,\footnote{%
 For two-sided Fourier transforms, 
 $ 
  C_{ab}(\omega) = [ 1/2
+ f^{ }_\rmii{B}(\omega) ] \rho_{ab}(\omega)
 $, 
with $\rho_{ab}$ from \eq\nr{rho}.
 } 
the one-sided Fourier transform can now be expressed as 
\begin{align} 
   C ^ + _ { a b } ( \omega  ) = 
   \int \! \frac{ {\rm d} \omega  ' } { 2 \pi  } \,  
   \frac{ i } { \omega  - \omega  ' } 
   \left [ \frac12  + f _ \rmii{B} ( \omega  ' ) \right ] 
   \rho  _ {  a  b } ( \omega  ' )
   \;, \label{short}
\end{align}
where $f _ \rmii{B} $ is the Bose distribution.  
Here the  
spectral function for bosonic operators $ X _ a $ and $ X _ b $ is
defined as
\begin{align} 
   \rho  _ { a b } ( \omega   ) \equiv 
   \int 
   \! {\rm d} t \, e ^{ i \omega  t } 
   \bigl\langle\,  [ X _ a ( t ) , X _ b ( 0 ) ]  \, \bigr\rangle 
   \label{rho}
   \;.
\end{align} 
If time reversal T is a symmetry, 
this spectral function is real (cf.\ appendix B of \Ref\cite{kadanoff}). 
If one neglects the CP violation in the Standard Model and 
in the Yukawa interactions of the RH neutrinos 
(the latter is of $O(h^4)$), 
CP is a symmetry, and so is~T. In this approximation
$ \rho  _ { a b }  $ is real. Then for real $\omega$,  
\begin{align} 
   { \rm Re } \, C ^ + _ { a b } ( \omega + i 0 ^ + ) = \frac12
   \left [ \frac12  + f _ \rmii{B} ( \omega   ) \right ] 
   \rho  _ { a  b } ( \omega  )
   \;. \label{cqft}
\end{align} 

Let us now compare eqs.~(\ref{ceff}) and (\ref{cqft}). 
We are interested in  $ \omega  \ll \omega^{ }_\rmii{UV} \,\lsim\, T $, 
in which case
we can approximate the expression 
in square brackets by $ T/\omega  $.
Matching the two expressions gives
\begin{align}
   \gamma  _ { a b } = \frac1{2V} \; \omega  
    \rho  _ { a c } 
    ( \omega  )
   \left ( \Xi  ^{ -1 } \right ) _ { cb }
   \;, \qquad  
   \mbox{for} \; 
   \gamma  \ll \omega  \ll \omega  _ \rmii{UV} 
   \;. \label{washout3.9.0a} 
\end{align} 
Here $ V $ is the spatial volume, 
and $ \Xi  $ is a matrix of susceptibilities, 
\begin{align} 
  \Xi  _ { a b } \equiv \frac{ 1 } { T V } \langle X _ a X _ b \rangle 
  \;. \label{sus}
\end{align}
As will be seen later on, $\Xi  _ { a b }$ is finite for $V\to\infty$, 
and the same is true for $\gamma _ {a b}$.
A consistency check is provided by whether $\rho_{ac}$ indeed
has a $1/\omega$-tail at $\omega \ll \omega_\rmii{UV}$.

The restriction on $ \omega  $ in \eq(\ref{washout3.9.0a}) limits the 
accuracy at which the $ \gamma  _ { ab } $ can be defined. 
The $\gamma _ {ab} $  are of 
order $ h
^ 2 \Lambda  $, 
where $ h
$ is a RH neutrino Yukawa coupling to be defined below,
and 
$  \Lambda  = \mbox{max} \{\pi T, M _ I \}$.\footnote{%
 Cf.\ \eq\nr{c1} for the way that mass and thermal 
 scales can be compared with each other.}
A rough estimate gives a relative
accuracy $ \gamma  / \omega  _ \rmii{UV} \sim h
^ 2 $ modulo coupling
constants which characterize spectator processes. It is therefore probably
not meaningful
to calculate the $ \gamma  _ { ab } $ 
beyond leading order in $ \hnu^2$. Of course,
radiative corrections due to SM 
interactions can be computed and can be important.

%
\subsection{The susceptibility matrix\label{sec:Xi}}

In order to make use of \eq\nr{washout3.9.0a} we need to 
determine the susceptibility matrix
$ \Xi  $ defined in \eq\nr{sus}, which turns out to 
require a little care (cf.\ \Ref\cite{khlebnikov}). 
It is perhaps simplest to think of the problem in 
an ensemble in which the values of the strictly conserved charges 
$ X _ { \bar{ a } }  $
are zero and do not fluctuate.\footnote{%
 This is equivalent to  a ``grand canonical'' 
 ensemble in which the charges $ X _ { \bar{ a } } $ 
 do fluctuate but their expectation values are zero, 
 cf.\ \eq\nr{Xi}. 
 } 
We write the Hamilton operator as 
\begin{align} 
   H = H _ 0 + H _ {\rm \Yuk } 
\end{align} 
where $ H _ 0 $ describes all free  
particles as well as their $ X _ a $-conserving interactions 
and $ H _ {\rm \Yuk }  $ is the 
interaction which violates 
$ X _ a $-conservation. The  susceptibilities 
are equal-time correlation functions, for
which $ H _ {\rm \Yuk } $ is a small perturbation. Since we only consider 
the leading order in $ h
$, the susceptibilities can be computed
using $ H _ 0 $, which commutes with the $ X _ a $. In this approximation 
the $ X _ a $ are conserved, and the ordering of the 
operators in \eq(\ref{sus}) 
does not matter. 

To compute the susceptibilities 
start with a grand canonical partition function 
\begin{align} 
  \exp \left (  -   \Omega   /T \right ) 
  =  \mbox{tr}\, \exp \left [ 
   ( \mu  _ A X _ A  - H _ 0)/T \right ]
  \;, 
\end{align}  
with chemical potentials $ \mu  _ A $ for all charges 
$ X _ A \in \{ X _ a, X _ { \bar a } \}$.   Some of the 
$ X _ { \bar a }  $ may be gauge charges. In this case
the role of  $ \mu  _ { \bar a }  $ is played by the zero-momentum mode
of the time component of the gauge field \cite{khlebnikov}. 
The thermodynamic
potential $ \widetilde{\Omega}  $  corresponding 
to fixed $ X _ { \bar a } = - \partial 
\Omega / \partial \mu  _ { \bar a }  $ is given by the Legendre transform 
$ \widetilde{\Omega}  = \Omega 
+ \mu  _ { \bar a } X _ { \bar a }  $. 
We are interested in $ X _ { \bar a }  = 0 $, 
so we have
$ 
    \widetilde{\Omega  }  = \Omega  
$ 
with  the $ \mu  _ { \bar a }  $ determined by
\begin{align} 
  \frac{ \partial \Omega  } { \partial \mu  _ { \bar a } }  = 0
  \label{Y=0}
  \;.
\end{align}  
The required susceptibilities are 
\begin{align} 
   \Xi  _ { a b } = - \frac1V 
   \left . \frac{ \partial ^ 2 \widetilde{\Omega  }   }
   { \partial \mu  _ a \partial \mu  _ b }  \right | _ { { \mu }  = 0 } 
   = 
   - \frac1V 
   \frac{ \partial ^ 2    }
   { \partial \mu  _ a \partial \mu  _ b } 
   \left\{  
   \left. \Omega \right|_{ \partial \Omega / \partial {\mu_{\bar{a}}} = 0 }   
    \right \} _ { { \mu  }   = 0 } 
 \;. 
 \label{Xi}
\end{align}

In order to evaluate \eq\nr{Xi} 
it is sufficient to expand $ \Omega  $ to second order,
\begin{align} 
  \Omega  = \Omega  _ 0  - \frac{  V } { 2 } 
  \mu^{ }  _ A \chi^{ }  _ { AB } \mu^{ }  _ B 
  + O(\mu^4)
  \;, 
  \label{Omega} 
\end{align} 
with the grand canonical susceptibilities
\begin{align} 
   \chi  ^{ }_ { AB }  \equiv 
   \frac{1}{TV} \langle X _ A X _ B \rangle _ {\rm grand \, canonical }
   = 
    - \frac1V 
   \left . \frac{ \partial ^ 2 {\Omega  }   }
   { \partial \mu  _ A \partial \mu  _ B }  \right | _ { { \mu }    = 0 } 
  \;. \label{susc}
\end{align} 
Then \eq(\ref{Y=0}) reads
\begin{align}
  \chi  _ { \bar{ a }\,  \bar{ b } }\,  \mu  _ { \bar{ b } } 
  = - \chi  _ { \bar{ a } \, b  } \mu  _ { b } 
  \label{washout11} 
  \;.
\end{align}
Here we see which strictly conserved charges need to be included: the ones 
which are correlated with strictly conserved charges which are correlated
with one of the $ X _ a $. The remaining ones
need not be taken into consideration. This is the case, e.g., 
for non-abelian gauge charges in the symmetric phase.
Solving \eq\nr{washout11} gives
$
   \mu  _ { \bar a } = - 
   \left ( { \xi  } \,{}  ^{ -1 } \right )^{ } _ { \bar a \bar b }  
    \chi^{ } _ { \bar b c } \,\mu^{ }  _ c 
$ 
where the matrix $  \xi   $ is defined by 
\begin{align} 
    \xi  _ { \bar a \bar b } \equiv \chi  _ { \bar a \bar b }
  \;.
\end{align} 
Inserting this into \eq\nr{Xi}  we find
\begin{align} 
  \Xi  _ { a b } = \chi^{ }  _ { a b } - 
  \chi^{ }_{ a \bar a } 
  \bigl( { \xi  } \,{}  ^{ -1 } \bigr)^{ } _ { \bar a \bar b } 
  \,\chi^{ }_{ \bar b b }  
  \label{washout12} 
 \;. 
\end{align}

%
\section{Lepton number washout rate}
\label{sec:washout}

%
\subsection{Master formula}

So far the discussion was general and did not make any use of
the specifics of the interaction which breaks the $ X _ a $-symmetry.
In the basic leptogenesis scenario 
the SM is extended by adding right-handed neutrino fields 
$ N _ {I } $ with Majorana masses $ M _ I $ 
(we employ a basis in which the Majorana mass matrix is diagonal). 
In the simplest realization they interact with the SM
particles via a Yukawa coupling to ordinary, left-handed lepton doublets 
$\ell_i \equiv \ell^{ }_{\rmii{L}i}$ 
and the Higgs doublet $ \varphi $ as follows:
\begin{align}\label{Lint}
   \mathcal{L}_{\rm int} = 
  {}-   \overline{N} \, h 
  \, {\widetilde \varphi}^\dagger \,
   \ell 
   + \mbox{h.c. }
   \;.
\end{align}
Here $\widetilde \varphi \equiv i \sigma ^ 2 \varphi
^ \ast $ with the Pauli matrix $ \sigma ^ 2 $ is the isospin conjugate
of $ \varphi $, and  the Yukawa couplings are written as a matrix in 
\flav space, $  \hnu  
= (h _ { I j } ) $.

In the simplest case it is only the interaction in \eq(\ref{Lint}) 
which is responsible for inducing slowly evolving processes. 
In more complicated cases some
SM Yukawa interactions proceed at a similar rate, 
and these interactions have to be 
taken into account as well~\cite{beneke}. 
Here we restrict ourselves to the first situation. 
What the relevant conserved and quasi-conserved charges $ X _ { \bar a }  $ 
and $ X _ a $ are, depends on the expansion rate of the Universe 
and thus on the 
temperature. One has to take into account all interactions which are much
faster than the expansion. 
The quasi-conserved charges
may contain many different fields, for instance 
both lepton and quark fields. For this
reason it is convenient 
to compute the spectral function of the time 
derivatives $ \dot X _ a $ of the charges 
rather than of the charges directly, and then use
\begin{align} 
  \omega  ^ 2 
   \rho  _ { a b } ( \omega )  
  = 
   \int 
  \! {\rm d} t\,  e ^{ i \omega  t } 
  \left \langle \left [ \dot X _ a ( t ) , \dot X _ b ( 0 ) \right ]
  \right \rangle 
  \;. \label{omega2rho}
\end{align} 
The operators $ \dot X _ a $ 
only contain the fields which interact via $ \mathcal{L} _ {\rm \Yuk } $. 
Furthermore, $ \dot X _ a $ contains $ \hnu $ explicitly. 
Since we compute only to leading order
in $ \hnu $, the thermal average on the 
right-hand side of \eq(\ref{omega2rho}) can be taken in
an ensemble with $ \hnu = 0 $. 
Therefore we can re-express \eq\nr{washout3.9.0a} as  
\begin{align}
   \gamma  _ { a b } = \frac1{2V} 
   \lim _ { \omega  \to 0 } \frac1\omega  
   \int 
   \! {\rm d} t\,  e ^{ i \omega  t } 
   \Big \langle \Big [ \dot X^{ } _ a ( t ) , \dot X^{ } _ c ( 0 ) \Big] 
   \Big \rangle^{ } _ 0
   \left ( \Xi  ^{ -1 } \right )^{ } _ { cb }
   \;, \label{washout3.9.0b}
\end{align} 
where the subscript $ 0 $ indicates that 
$ h
= 0 $ in the average,
so that $ h
$ only appears in the operators. 
Equation~(\ref{washout3.9.0b}) has some similarity 
with Kubo formulas for transport coefficients~\cite{kubo,kadanoff}, 
in particular for \flav diffusion (cf.\ \Ref\cite{arnold-ll}). 

As already mentioned, the choice of the broken charges is not unique: 
adding some linear combination
of the conserved ones we again obtain a charge which is not conserved. 
It is possible to choose the symmetries so that they only act on SM particles. 
Let the left-handed leptons transform as 
\begin{align} 
   \ell^{ } _ i \to \left ( e ^{ i \alpha  _ a T _ a ^ \ell} \right )
   _ { ij } \ell^{ } _ j
  \;, 
  \label{Ta}
\end{align} 
with Hermitian matrices  $ T _ a ^ \ell$.
The interaction Lagrangian in \eq(\ref{Lint}) 
is not invariant under this transformation.
Following the  usual steps  to derive Noether's theorem one finds
\begin{align} 
   X _ a = \int _ \vec x 
   \, 
  \Bigl[ 
   \sum_i  
  \overline{ \ell\,}_{\! i}  \gamma  ^ 0 T _ a ^ \ell \ell_i  
  + (\mbox{contributions from other fields})
  \Bigr]
  \;, \label{Xa_def}
\end{align} 
and
\begin{align}
   \dot X _ a = i \int  _ \vec x 
   \Big [ \overline{ N }\, \hnu\, 
   \widetilde{\varphi  } ^\dagger
   \,T _ a ^ \ell \, \ell - 
   \overline{ \ell\, }\, T _ a ^ \ell\, \widetilde{\varphi  }\, 
   \hnu ^\dagger
   N \Big ]
   \;. \label{washout31.1}
\end{align} 

Given that the thermal average 
in \eq\nr{washout3.9.0b}
is performed with $ \hnu  
= 0 $, we can integrate out the
RH neutrinos treating them as free fields. 
In a basis where the Majorana mass matrix is diagonal, 
a straightforward calculation\footnote{%
 For instance, making use of the imaginary-time formalism, 
 \nr{washout3.9.0b} contains
 a 2-point correlator of the right-handed neutrino fields and of the composite
 operators to which they couple according to 
 \nr{washout31.1}. 
 The former is of the familiar
 form $(\slash\!\!\!{k} + M_I) /(k^2 - M_I^2)$, with $k^0 \to i k_n$; 
 the mass in the 
 numerator is projected out by the Dirac trace. The latter can be expressed
 in a spectral representation as $\Sigma(p_n,{\bf p}) = 
 \int \! {\rm d}p^0/(2\pi)\,  \widetilde{\rho}(p^0,{\bf p}) /
 (p^0 - i p_n)$, where $\widetilde{\rho}$ is from \nr{rho_dX}. Matsubara sums 
 can be carried out, and generate Fermi-Dirac distributions.  
 A subsequent analytic continuation yields a retarded
 real-time correlator, and its cut yields 
 the spectral function needed in \nr{washout3.9.0b}.
 This contains structures like 
 $f^{ }_\rmii{F}(E_I + \omega) - f^{ }_\rmii{F}(E_I)$, which after 
 taking the limit $\lim_{\omega\to 0} (...)/\omega$ 
 leave over $f'_\rmii{F}(E_I)$.
 }
yields for \eq\nr{washout3.9.0b} 
\begin{align} 
  \gamma    _ { ab } 
  = 
  &
  - \frac12   \sum _ I 
   \int _ \vec k \frac{ f' _ \rmii{F} ( E _ I )   } { 2 E _ I    }
   \nonumber \\
   \times &
   \, h _ { I i } 
   \mbox{   tr} \left  [\, \cancel{k } \, 
   \Big ( T _ a ^\ell \big [ \, 
   \widetilde{ \rho } ( k ) + \widetilde{ \rho  } ( -k ) \,
   \big ] T _ c ^ \ell
         + T _ c ^ \ell \big[ \,
   \widetilde{ \rho } ( k ) + \widetilde{ \rho  } ( -k )
   \, \big] T _ a ^\ell 
   \Big ) _ { ij } \right  ] h ^ \ast _ { Ij } 
   \left ( \Xi  ^{ -1 } \right ) _ { cb }  
   \;, \label{washout44.3}
\end{align} 
where the trace refers to the spinor indices, and 
$ k ^ 0 =  E _ I \equiv ( \vec k ^ 2 + M _ I ^ 2 ) ^{ 1/2 }  $. 
The prime is a derivative 
with respect to energy, and  $ f _ \rmii{F} $ is 
the Fermi-Dirac distribution.
We have introduced the spectral function for the composite
operator of the SM fields 
that appears in \eq(\ref{washout31.1}), 
\begin{align} 
   \widetilde{ \rho }  _ { ij \alpha  \beta  } ( k ) 
   \, \equiv \int _ x 
   e ^{ i k \cdot x } 
   \Bigl\langle \, 
      \Bigl\{ 
   (\widetilde{\varphi  } ^\dagger \ell _ { i \alpha  }) ( x ) , 
   (\overline{ \ell\, } _ {\! j \beta  }\, \widetilde{\varphi  }) ( 0 )  
      \Bigr\}
   \, \Bigr \rangle ^{ }_ 0
   \;, \label{rho_dX}
\end{align} 
where $ \alpha  $, $ \beta  $ are Dirac spinor indices. 
Note that for fermionic operators the spectral function 
is defined with anticommutators.

Equations \nr{washout44.3}, \nr{rho_dX}, together with the expression
for the $\Xi$ matrix in \eq\nr{Xi} for charges like those 
in \eq\nr{Xa_def} 
constitute the main formal results of this paper. We stress again 
that these expressions are valid to any order in SM couplings.   

If charged lepton Yukawa interactions 
can be neglected, $ \widetilde{ \rho  } $ 
is invariant under $ U ( 3 ) _ \ell $  and thus
$ \widetilde{ \rho  } _ { ij }  \propto \delta  _ { ij } $. 
Writing 
$
   \widetilde{ \rho  } _ {  i j  \alpha  \beta} ( k ) 
 = 
  \delta_{ij} \,  
  \widetilde{ \rho  } _ { \alpha  \beta  } ( k ) 
$, 
we can then re-express \eq\nr{washout44.3} as 
\ba
    \gamma  _ { ab } & = & \fr12 \sum_{I} 
    h _ { I i } \{ T _ a ^\ell , T _ c ^ \ell \}^{ } _ { ij }
    h ^ \ast _ { Ij } 
   \left ( \Xi  ^{ -1 } \right ) _ { cb } 
   \, \mathcal{W}(M_I)
  \;, \label{pre_washout_rate} \\ 
  \mathcal{W}(M_I)
  & \equiv &  -  
  \int _ \vec k\frac{ f _ \rmii{F}' ( E _ I )  } { 2 E _ I   }
  \, 
   ( \cancel{ k } )_{\beta\alpha}
  \big [ 
   \widetilde{ \rho }_{\alpha\beta} ( k )
 + \widetilde{ \rho  }_{\alpha\beta} ( -k )\big ]
 \;. \label{washout_rate}
\ea

\subsection{Spectral function}
\label{sec:spectral}

We proceed to discussing how the real-time part of the washout rate, 
i.e.\ the weighted integral over $\widetilde{ \rho }$  
in \eqs\nr{washout44.3} and \nr{washout_rate}, 
can be evaluated in practice. 

%
\subsubsection{Leading order in the non-relativistic regime}
\label{ss:rho_lo}

Consider first the LO contributions 
to $ \widetilde{ \rho } $ in the symmetric phase 
in a regime  $ M _ I \gg \pi T$. 
In this case one can neglect the thermal masses of the SM
particles\footnote{%
 Thermal masses have to be taken into account 
 for $ M  _ I \lsim \sqrt{ g } T $~\cite{relat}.}
which can then be treated as massless and free. 
For timelike $ k $ 
($ k ^ 2 > 0 $) but with $k^0$ of either sign one obtains 
\begin{align}
   \widetilde{ \rho  } _ {  \alpha  \beta } ( k ) 
   = 2 
   \int _ \vec p \frac{ 1 } { 2 p ^ 0  }  
   &
   \left ( P _ \rmii{L}\, \cancel{ p } \right ) _ { \alpha  \beta  } 
   \int _ \vec q \frac{ 1 } { 2 q ^ 0  }  
   \big [ 1
      - f _ \rmii{F}\! \left ( p ^ 0 \right ) 
      + f _ \rmii{B}\! \left ( q ^ 0 \right ) 
   \big ]
   \nonumber \\
   &
   \times ( 2 \pi  ) ^ 4 
   \Big[ \delta^{4}  ( p + q - k ) + \delta^{4}  ( p + q + k) \Big ] 
  \;, 
\end{align} 
where $ p ^ 0 \equiv | \vec p | $, $ q ^ 0 \equiv | \vec q | $, 
and the left chiral projector is defined as $ P _ \rmii{L} \equiv
( 1 - \gamma   _ 5 ) /2 $. 
For 
\eq\nr{washout_rate} we get
\begin{align} 
  \mathcal{W}(M_I) = & - 4
   \int _  \vec k \frac{ 1 } { 2 E _ I    }
 \int _ \vec p \frac{1 } { 2 p ^ 0  }  
 \int _ \vec q \frac{ 1 } { 2 q ^ 0  }  
   \nonumber \\
   & \quad \times \, 
   2 p \cdot k 
   \, f' _ \rmii{F} ( E _ I ) \, 
   \big [ 1
      - f _ \rmii{F}\! \left ( p ^ 0 \right ) 
      + f _ \rmii{B}\! \left ( q ^ 0 \right ) 
   \big ]
   ( 2 \pi  ) ^ 4 \delta^4  ( p + q - k ) 
   \;. \label{washout64}
\end{align} 
When  $ M _ I \gg \pi T $, the 
Bose and Fermi distributions
in the square brackets 
are exponentially suppressed 
with $ \exp (  { - M _ I / T  } ) $ 
and can be neglected. 
Omitting terms of  order $ \exp (  { - M _ I / T  } ) $
also in $f'_\rmii{F}(E_I)$, it is straightforward to perform the 
integrals, yielding
\begin{align} 
   \gamma  _ { ab } = \frac{ 1 } { 16 \pi  ^ 3 }
   \sum _ I M _ I ^ 3 \, K^{ } _ 1 
   (  M _ I / T ) \,
   h   _ { I i } \{ T _ a ^\ell , T _ c ^ \ell \} ^{ }_ { ij }
     h ^ \ast _ { Ij } 
   \left ( \Xi  ^{ -1 } \right ) _ { cb } 
   \;, \quad  \pi T \ll M _ I  
   \;. \label{washout67}
\end{align} 
Here $ K _ 1 $ is a modified Bessel function of the second kind.
Once a LO susceptibility from \se\ref{ss:susc_lo}
is inserted, 
this reduces to a standard result, cf.\ \eq\nr{standard}. 

%
\subsubsection{Next-to-leading order}

In general  the spectral function in \eq\nr{rho_dX} contains 
two independent Dirac structures
at finite temperature~\cite{haw}. However the 
Dirac trace in \eq\nr{washout44.3} is exactly the same as appears
in the right-handed neutrino production rate, projecting
out a particular 
linear combination of the Dirac structures. Assuming as before that 
$
   \widetilde{ \rho  } _ {  i j  \alpha  \beta} ( k ) 
 = 
  \delta_{ij} \,  
  \widetilde{ \rho  } _ { \alpha  \beta  } ( k ) 
$, 
the production rate of flavour $I$ reads 
\ba
  \gamma_I^+  & = &   
  \sum_i |h^{ }_{Ii}|^2
  \, \mathcal{P}(M_I)
  \;, \\ 
  \mathcal{P}(M_I)
  & \equiv &   
  \int _ \vec k\frac{ f _ \rmii{F} ( E _ I ) } { 2 E _ I   }
   \,
   ( \cancel{ k } )_{\beta\alpha}
  \big [ 
   \widetilde{ \rho }_{\alpha\beta} ( k )
 + \widetilde{ \rho  }_{\alpha\beta} ( -k )\big ]
 \;. \label{production_rate}
\ea
Comparing with \eq\nr{washout_rate}, it is seen that 
$\mathcal{W}$ differs from $\mathcal{P}$ only through  
a weight, $- f' _ \rmii{F}(E_I)$ versus $f _ \rmii{F}(E_I)$.\footnote{%
 An intuitive reason for the difference is that in the production rate 
 the combination $\sim f _ \rmii{F}(E_I+\mu) + f _ \rmii{F}(E_I-\mu)$ 
 appears 
 whereas in the dissipation rate it is the difference 
 $\sim f _ \rmii{F}(E_I+\mu) - f _ \rmii{F}(E_I-\mu)$ that plays a role. 
 Here $\mu$ is a chemical potential induced by the Yukawa interaction. 
 } 
Therefore, $\mathcal{W}$ can be extracted from known results 
for $\mathcal{P}$.

\begin{figure}[t]


\centerline{%
\epsfysize=9.5cm\epsfbox{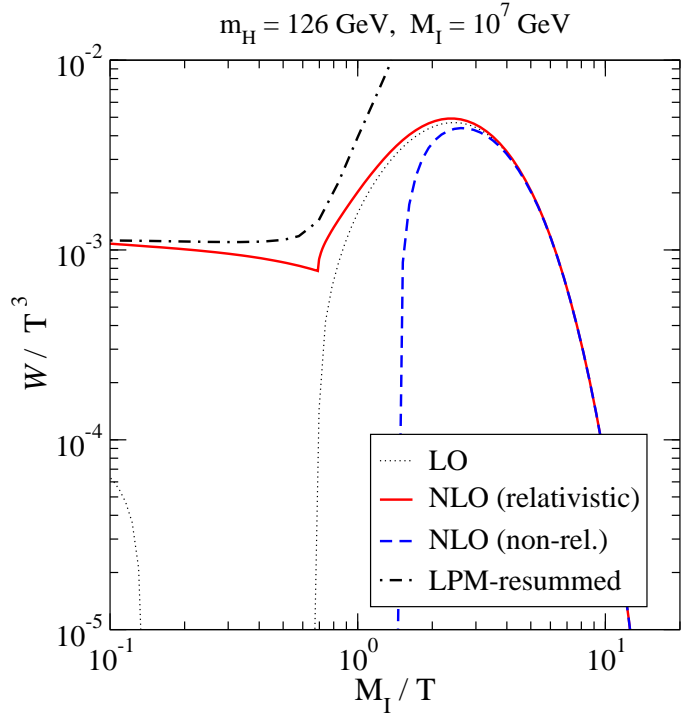}%
}

\caption[a]{\small
The washout rate expressed through $\mathcal{W}$
as defined in \eq\nr{washout_rate}. 
Shown are the LO result (dotted line); 
the relativistic NLO result (solid line, from ref.~\cite{relat});  
the NLO result in the non-relativistic approximation (dashed line, 
from eq.~\nr{washout_nlo}); 
and the LPM-resummed result valid in 
the ultrarelativistic regime $M_I \lsim g T$ 
(dash-dotted line, from ref.~\cite{besak}). 
}

\la{fig:washout}
\end{figure}

The extraction of $\mathcal{W}$ is particularly simple
in the non-relativistic regime $\pi T \ll M_I$, where
$f' _ \rmii{F}(E_I) = - f _ \rmii{F}(E_I)/T$. 
Displaying contributions involving
the U$^{ }_
{Y}$(1), SU$^{ }_\rmii{L}$(2) and SU(3) 
gauge couplings $g_1,g_2$ and $g_3$, the Higgs self-coupling $\lambda$, 
with the tree-level value $\lambda = g_2^2 m_H^2 / (8m_W^2)$, 
as well as the top Yukawa coupling $h_t$, the Dirac trace
reads~\cite{nonrel}
\be
 ( \cancel{ k } )_{\beta\alpha}
 \big [ 
   \widetilde{ \rho }_{\alpha\beta} ( k )
 + \widetilde{ \rho  }_{\alpha\beta} ( -k )\big ]
 = \frac{ M_I^2 }{2\pi}
 \left [ 1 + c_1 +  \frac{c_2 \, \vec{k}^2 }{M_I^2}  + 
 O \left ( \frac{ \vec{k}^4 }{ M_I^4 } \right ) \right ]
 \;,
\ee
where 
\ba
 c_1 & = & 
  - \frac{\lambda T^2}{M_I^2} \biggl( 1 - \frac{3m_\varphi}{\pi T} \biggr) 
   \, - \, 
   |h_t|^2 
   \biggl[ \frac{3}{(4\pi)^2}
    \biggl( \ln\frac{\bar{\mu}^2}{M_I^2} + \frac{7}{2} \biggr) + 
     \frac{7\pi^2 T^4 }{60 M_I^4}
   \biggr]
    \nonumber \\ 
   & & \qquad + \, 
   (g_1^2 + 3 g_2^2)
   \biggl[ \frac{3}{4(4\pi)^2} \biggl( 
     \ln\frac{\bar{\mu}^2}{M_I^2} + \frac{29}{6} \biggr) - 
     \frac{\pi^2 T^4}{80M_I^4}
   \biggr] +
  O\Bigl(g^4,\frac{g^2T^6}{M_I^6} \Bigr)
 \;, \label{c1} \\ 
 c_2 & = & 
   \, - \, 
   |h_t|^2 
      \frac{7\pi^2 T^4 }{45 M_I^4}
     \, - \, 
   (g_1^2 + 3 g_2^2)
     \frac{\pi^2 T^4}{60M_I^4} 
  + O\Bigl(\frac{g^4T^4}{M_I^4},\frac{g^2T^6}{M_I^6} \Bigr) 
 \;. \label{c2} 
\ea
Here $\bar{\mu}$ is the $\msbar$ renormalization scale related to 
the neutrino Yukawa couplings appearing in \eq\nr{pre_washout_rate},
and $m_\varphi$ is the thermal Higgs mass parameter, 
\be
  m_\varphi^2 \,\equiv\,  m_0^2 +   \frac{T^2}{16}
 \Bigl(
   g_1^2 + 3g_2^2 + 4 |h_t|^2  + 8 \lambda 
 \Bigr)
 \;, \label{m_phi}
\ee
where $m_0^2 < 0$ is the vacuum value, and we assume 
$m_\varphi^2 > 0$.
Integrating over the momenta in \eq\nr{washout_rate} yields
corrections to \eq\nr{washout67}: 
\ba
   \gamma  _ { ab } & = &  \frac{ 1 } { 16 \pi  ^ 3 }
   \sum _ I M _ I ^ 3 \, 
   \biggl[ (1 + c_1)\, K^{ } _ 1 \left ( \frac{ M _ I } T \right ) + 
   \frac{3 c_2 T}{M_I} K^{ } _ 2 \left ( \frac{ M _ I } T \right )
   + O \Bigl( \frac{T^{17/2} e^{-M_I/T}}{M_I^{17/2}}  \Bigr)
   \biggr] 
   \nonumber \\ 
   & & \qquad \times \,   
   h _ { I i } \{ T _ a ^\ell , T _ c ^ \ell \}^{ } _ { ij }
     h ^ \ast _ { Ij } 
   \left ( \Xi  ^{ -1 } \right ) _ { cb } 
   \;, \quad  \pi T \ll M _ I  
   \;. \label{washout_nlo}
\ea

If the temperature is increased to 
$T \gsim M_I/4$, which may be relevant e.g.\ 
for setting the initial conditions for leptogenesis, 
we leave the non-relativistic regime~\cite{relat}.
The NLO  production rate 
in the relativistic regime ($\pi T \sim M_I$) and the LO
production rate
in the ultrarelativistic regime ($g T \gsim M_I$) 
are also known but only in numerical form~\cite{relat,besak}.\footnote{%
 For $g T  \gsim  M _ I$, 
 multiple gauge interactions need to be resummed to obtain
 the correct LO result; the resummation can be expressed 
 as a solution of an inhomogeneous 
 differential equation~\cite{anisimov,besak}.}
The results of refs.~\cite{relat,besak} 
for $\mathcal{P}(M_I) / T^4$ are plotted in fig.~5 of ref.~\cite{relat}. 
The results of refs.~\cite{relat,besak} for $\mathcal{W}(M_I) / T^3$, 
obtained by changing the weight from 
$f _ \rmii{F}(E_I)$ to $- f' _ \rmii{F}(E_I)$, 
are shown in \fig\ref{fig:washout}.
It is seen how the perturbative expansion breaks down and 
resummations are necessary for $T \gsim M_I$, and how the subsequent
washout rate is strongly enhanced compared with 
a naive tree-level analysis (``LO'' in the plot). 
It is also clear that the non-relativistic expansion
of \eq\nr{washout_nlo} breaks down for $T \gsim M_I / 4$.
(To be more precise, the non-relativistic expansion shows convergence
only for $T \lsim M_I / 15$, but the smallness of {\em any} loop 
corrections allows it to be used in practice 
up to somewhat higher temperatures~\cite{relat}.)

%
\subsection{Susceptibility matrix}
\label{se:suscs}

The  susceptibilities as given by \eq\nr{sus}
or \eq\nr{Xi} measure the correlations in fluctuations of the slowly varying
charges $X_a$ when the other charges  $X_{\bar{a}}^{ }$
are constrained to vanish. The susceptibility matrix influences the 
lepton number washout rate as indicated 
in \eq\nr{washout3.9.0a}, 
\nr{washout44.3} and \nr{pre_washout_rate}.  
Here we show how the susceptibilities 
can be computed in practice, first at leading
order and then including corrections of $O(g)$ and $O(g^2)$.
We consider the Standard Model with left-handed quarks 
($q \equiv q^{ }_\rmii{L}$) and leptons ($\ell \equiv \ell^{ }_\rmii{L}$); 
as well as right-handed up-type quarks ($u \equiv u^{ }_\rmii{R}$), down-type
quarks ($d \equiv d^{ }_\rmii{R}$), and charged
leptons ($e \equiv e^{ }_\rmii{R}$).

%
\subsubsection{Leading order}
\label{ss:susc_lo}

The susceptibilities defined by \eq\nr{
sus} are,  at LO, 
determined by free field theory.
It turns out that at NLO it is helpful to first evaluate the grand canonical 
potential $\Omega$, and then extract the answer directly from the second 
relation in \eq\nr{Xi}. However, we start by showing
how at LO the results can also be obtained from the
2-point correlators in \eq\nr{susc} through \eq\nr{washout12}.  

For \eq\nr{susc} one has to compute fluctuations
of charges, which for free fields are 
in one-to-one correspondence with particle number
fluctuations. Start with the  fluctuation $ \langle Q ^ 2 \rangle $ 
where $ Q $ is the difference of particle and antiparticle number  
of a single fermion species. 
By a single fermion species we mean either one chiral 
fermion or a single spin state of a Dirac fermion.
For only one fermionic degree of freedom, 
we have the 
particle number fluctuation 
$ \langle ( \Delta  N ) ^ 2 \rangle = V T ^ 3 /12 $. The fluctuations
of particles and antiparticles are uncorrelated which 
implies $ \langle Q ^ 2 \rangle = V T ^ 3 /6 $. 

The fluctuations of the charges 
\begin{align} 
   Q _ a =  
   \int _ \vec x 
   \overline{ \psi } \gamma  ^ 0  T _ a \psi  
   = 
   \int _ \vec x 
   \psi  ^\dagger T _ a \psi  
\end{align} 
of a  set of left-chiral fermion fields $ \psi _ i $, 
$ \psi  _ i  = P _ \rmii{L} \psi _ i  $,
are given by 
\begin{align} 
   \langle Q _ a Q _ b \rangle = V \int _ \vec x 
   \left \langle 
   \left ( \psi  ^\dagger  T _ a \psi \right ) ( x ) 
   \left ( \psi  ^\dagger  T _ b \psi \right ) ( 0 ) 
   \right \rangle 
   \;. 
   \label{mu13.1.3}
\end{align} 
The free propagators are \flav diagonal which directly implies
$
   \langle Q _ a Q _ b \rangle = V T \chi  _ { ab } 
$ 
with the susceptibilities 
\begin{align} 
   \chi  _ { ab }  = 
   \mbox{ tr} \left ( T _ a T _ b \right  ) \frac{ T ^ 2 }6
  \;. \label{chi_ab}
\end{align} 
We also need the fluctuation of the weak hypercharge $ Y_\varphi $   
of the Higgs field. 
For a single scalar particle species (including the antiparticle) of unit
charge the charge fluctuation is $ \langle Q ^ 2 \rangle = V T ^ 3/3 $. 
Therefore, counting both isospin states,  
\begin{align} 
   \langle Y _ \varphi  ^ 2 \rangle = V T 
  \times 
   \frac23 y _ \varphi  ^ 2 T ^ 2 
 \;,
\end{align} 
with the Higgs hypercharge $ y _ { \varphi  } = 1/2 $.

Consider now the simplest realistic case, with only
one right-handed neutrino $ N _ 1 $ at a temperature
$ T \gg 10 ^ { 13 } $ GeV. Then
only top Yukawa and gauge interactions are in equilibrium. All  
other SM Yukawa interactions, as well
as strong and weak sphalerons
can be neglected. It is then sufficient
to consider only the left-handed
leptons $\ell$, the Higgs, 
the 3rd family quark doublet $ q _ 3 $ and 
the right-handed top $ t $. Without $ H _ {\rm int } $ we have a
U(3)$ {}_ \ell $ symmetry as well as two U(1) symmetries, generated
by the baryon number 
$ B _ { q _ 3 t } $ carried by $ q _ 3 $ and $ t $, and by the 
 hypercharge $ Y _ {  q _ 3 t \varphi  } $ carried by $ q _ 3 $, $ t $, and
$ \varphi  $ 
(we denote $ X _ { AB \,\cdots  } \equiv X _ A + X _ B  
+ \cdots  $).\footnote{%
  In principle one could have chosen the total 
  hypercharge $ Y $ as one of the
  $ X _ { \bar{ a } } $. However, this would be rather inconvenient because
  there are a lot of conserved charges which are 
  correlated with $ Y $ and which would then 
  all have to be included in the set of $ X _ A $.
  Note also that when employing \nr{chi_ab} 
  one has to keep in mind that the $ T_a $ may contain unit
  matrices in colour or weak isospin space which would contribute factors
  of $N_{\rm colour}$ or $N_{\rm weak\, isospin}$ to the trace 
  (cf.\ sec.~\ref{s:Xi-nlo}).
} 
Using a U(3)$ {}_ \ell $  transformation we choose
the fields such that $ N _ 1 $ only couples to the 
$ \ell $ of one family, which we denote by $ \ell _ { N _ 1 }$. 
Without $ H _ {\rm int} $ the corresponding lepton number 
$ L _ {  N _ 1  } $ is conserved.
It is broken by $ H _ {\rm int} $, together with  
$ Y _ {  q _ 3 t \varphi  } $, leaving 
$ Y _ {  q _ 3 t \varphi  \ell _ {\! N _ 1 } } $
unbroken. 
Thus the only $ X _ a $ can be chosen as  $ L _ {  { N _ 1 } } $,
and the set of $ X _ { \bar{ a } } $ consists of 
$ Y _ {  q _ 3 t \varphi  \ell _ {\! N _ 1 } } $ and 
$ B _ { q _ 3 t } $.\footnote{%
  Without $ H _ {\rm int} $ the 
  charges corresponding to the off-diagonal generators 
  which mix $ \ell _ { N _ 1 } $ with the other families are conserved 
  as well, and they are also broken by $ H _ {\rm int} $.  Here we consider
  only the dissipation of $ L _ {  N _ 1  } $; the evolution of 
  diagonal and off-diagonal charges decouples at leading order.}
If we  arrange the charges in this order we obtain  at LO 
\be 
   \chi  = 
   \left ( 
   \begin{array}{ccc}
     1/3     &  - 1/6   &   0 \\
     - 1/6   &  1/2    & 1/6 \\
    0            & 1/6      & 1/6    
   \end{array}
   \right ) 
   T ^ 2 
  \;. \label{X_ex0}
\ee 
Then from \eq\nr{washout12} $ \Xi  $ is just a number, 
$   \Xi  = T ^ 2 /4  $. In this scenario $ 
h _ { I i } \{ T _ a , T _ c \}^{ } _ { ij }
     h ^ \ast _ { Ij } = 2  | h _ {  11 } |^2  $. 
Undoing the U(3)$ {}_ \ell $ rotation, eq.~(\ref{washout67})
subsequently  gives the LO result  \cite{nrlg}
(it corresponds to $ c _ \ell = 1 $, $ c _ \varphi  = 2/3 $
in the notation of ref.~\cite{davidson-review})
\be  
   \gamma^{ }_{L_{N_1}}  
   = \sum_i \frac{ | h _ {  1i } | ^ 2  } 
   { 2 \pi  ^ 3 } \frac{ M _ 1 ^ 3 } { T ^ 2 } 
   \, K^{ } _ 1 \left ( \frac{ M _ 1 } T \right ) 
 \;. \label{standard}
\ee

%
\subsubsection{Next-to-leading order} 
\label{s:Xi-nlo}

We now include Standard Model interactions. 
The up-type, 
down-type, and charged lepton Yukawa couplings are 
denoted by $h^{ }_{uij}$, 
$h^{ }_{dij}$, $h^{ }_{eij}$, respectively, 
where $i,j \in \{1,2,3\}$ 
label families. 

In order to determine the susceptibilities, it is convenient 
to first compute the pressure~\cite{av}, $P(T,
\mu 
)$, 
as a function of the temperature and 
the chemical potentials associated with all
charges $ X _ A $. The pressure determines the
grand canonical potential through 
\begin{align} 
   \Omega = - P(T,
 \mu  ) V
 \;.
\end{align}  
One should include only  interactions which are in thermal equilibrium
(see the general discussion in \ses\ref{sec:physical} and~\ref{s:charges}). 
We assume this is the case for  the gauge interactions, and, depending
on the Hubble rate and thus on the temperature, 
some Standard Model Yukawa interactions. 

If one collects all fermion fields in one big spinor $ \psi  $, the 
fermionic contribution to the conserved charges can be written as 
\begin{align} 
   X _ A = \int  _ \vec x 
   \overline{ \psi  } \gamma  ^ 0 T _ A \psi  
 \;, 
\end{align} 
with hermitian matrices $ T _ A $.  
Including the chemical potentials
corresponds to  adding a term $ \overline{ \psi  } \gamma  ^ 0 \mu   \psi  $ 
with
$ 
   \mu  \equiv \mu  _ A T _ A 
$
to the Lagrangian.  For simplicity we carry out the computation 
in the symmetric phase of the electroweak theory. In this
situation the
$ T _ A $ commute with weak isospin rotations in addition
to \col rotations.  They are thus block diagonal and can be written as
$ T _ A = T _ A ^ q \otimes  \mathds{1} _ {\rm \col } \!
\otimes  \mathds{1} _ {\rm weak \, isospin } 
+ T _ A ^  u \otimes  \mathds{1} _ {\rm \col } \! + \cdots  $. 
Correspondingly,
the chemical potential matrix takes the form 
$ \mu  = \mu  _ q \otimes  \mathds{1} _ {\rm \col } \!
\otimes  \mathds{1} _ {\rm weak \, isospin } 
+  \mu  _ u \otimes  \mathds{1} _ {\rm \col } \! + \cdots  $, 
where $ \mu  _ q $, $ \mu  _ u , \ldots  $ are matrices in 
family space. 
Then the fermion propagators are matrices in family space as well. 

The dependence of $P(T,
\mu  )$ on the chemical potentials is needed
only up to quadratic order (cf.\ \eq\nr{susc}). 
The results of sec.~\ref{ss:susc_lo} correspond to 
(unresummed) 1-loop contributions
to the pressure. The Standard Model interactions enter 
at two loops. 
All 1- and 2-loop Feynman diagrams 
are displayed in fig.~\ref{fig:graphs}.
Since gauge interactions
are \flav blind, the chemical potentials 
can be diagonalized. 
In the Lagrangian 
\begin{align}
   {\cal L } _ {\rm SM-Yukawa} = 
  - \overline{ u } \, h _ u \, \widetilde{ \varphi  }^\dagger \, q 
   - \overline{ d } \, h _ d \,\varphi^\dagger \, q 
   - \overline{ e } \, h _ e \, \varphi^\dagger \, \ell 
   + \mbox{ h.c. }
   \label{LYuk} 
   \;
\end{align}
we include only  Yukawa interactions which are in equilibrium.
The terms included have to be invariant under the
symmetry transformations generated by the $ X _ A $. 
This implies relations between the $ T _ A $ and
the  in-equilibrium Yukawa couplings, 
\begin{align} 
   - T _  { A } ^ {  u }  h _ u + h _ u T ^  \varphi _ {   A }
 + h _ u T _ { A }  ^ q  
 & = 0
   \;, 
   \nonumber 
   \\
   - T _  { A } ^ {  d }  h _ d - h _ d T ^  \varphi _ {   A }
 + h _ d T _ { A } ^ q 
 & =  0
   \;, 
   \nonumber 
   \\
   - T _  { A } ^ {  e }  h _ e - h _ e T ^  \varphi _ {   A }
  + h _ e T _ { A } ^ \ell & =  0
   \label{mu201.2} 
   \;, 
\end{align} 
where $ T ^  \varphi _ {   A } $ is simply a number. 
Multiplying by $ \mu  _ A $ 
one finds relations between the chemical potentials, 
\begin{eqnarray} 
   - \mu  _ u h _ u + h _ u (  \mu  _ q  + \mu  _ \varphi ) 
  & =& 0
   \;,\nonumber 
    \\
    - \mu  _ d h _ d + h _ d ( \mu  _ q  - \mu  _ \varphi ) 
   & = & 0
   \;,\nonumber 
   \\
   - \mu  _ e h _ e + h _ e (\mu  _ \ell  - \mu  _ \varphi )  
    & = & 0
    \label{murelation} 
   \;. \hspace*{5mm} 
\end{eqnarray} 
Yukawa couplings mediating reactions 
not in equilibrium have to be omitted. 

By making use of \eqs(\ref{murelation}) and 
their hermitian conjugates, as well as 
substitutions of sum-integration variables, the 2-loop 
computation can be reduced to products of the following 1-loop sum-integrals: 
\ba
 T\sum_{p_n} \int_\vec{p} 
 \, \frac{1}{(p_n - i \mu)^2 + \vec{p}^2 + m_\varphi^2 }
 & = & 
 \int_\vec{p} 
 \frac{1 + f^{ }_\rmii{B}(E_\varphi - \mu) +
 f^{ }_\rmii{B}(E_\varphi + \mu) }{2 E_\varphi}
 \nonumber \\ 
 & = & 
 \frac{T^2}{12}\biggl(1 -  \frac{3 m_\varphi}{\pi T}  \biggr) + 
 \frac{\mu^2}{8 \pi^2} \biggl( \frac{\pi T}{m_\varphi} -1 \biggr) 
 + \ldots 
 \;, \hspace*{5mm} \label{tadpole_b} \\ 
 T\sum_{\{p_n\}} 
 \int_\vec{p}
 \, \frac{1}{(p_n - i \mu)^2 + \vec{p}^2  }
 & = & 
 \int_\vec{p} 
 \frac{1 - f^{ }_\rmii{F}(|\vec{p}| - \mu) -
 f^{ }_\rmii{F}(|\vec{p}| + \mu) }{2 |\vec{p}|}
 \nonumber \\ 
 & = & 
 - \frac{T^2}{24} - 
 \frac{\mu^2}{8 \pi^2} 
 \;. \label{tadpole_f}
\ea
Here  $p_n$ denotes bosonic
and $\{p_n\}$ fermionic Matsubara frequencies.
The parameter $m_\varphi^2$ is the thermal Higgs 
mass given by \eq\nr{m_phi},    
$E_\varphi \equiv ( \vec{p}^2 + m_\varphi^2  ) ^{ 1/2 } $, 
and we assume $\mu^2 \ll m_\varphi^2  \ll (\pi T)^2$. 
The bosonic result in \eq\nr{tadpole_b} is an expansion with 
higher orders omitted, whereas 
the fermionic result in \eq\nr{tadpole_f} is exact
(in dimensional regularization). 
It is important to keep in mind that the divergent $1/m_\varphi$ terms, 
appearing through \nr{tadpole_b}, 
need to be ``daisy resummed'' (or ``thermal mass resummed'')
in order to obtain a consistent 
weak-coupling expansion \cite{carrington}. 

%
\begin{figure}[t]
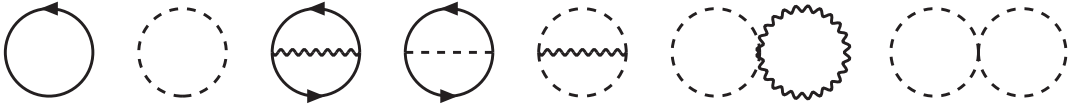


\begin{eqnarray*}
&& 
 \hspace*{-1cm}
 \TopoVR(\Aqu)
 \hspace*{0.8cm}
 \TopoVR(\Asc)
 \hspace*{0.8cm}
 \ToptVS(\Aqu,\Aqu,\Lgl)
 \hspace*{0.8cm}
 \ToptVS(\Aqu,\Aqu,\Lsc)
 \hspace*{0.8cm}
 \ToptVS(\Asc,\Asc,\Lgl)
 \hspace*{0.8cm}
 \ToptVE(\Asc,\Agl)
 \hspace*{1.2cm}
 \ToptVE(\Asc,\Asc)
\end{eqnarray*}

\caption[a]{\small 
1- and 2-loop graphs contributing to the pressure $P(T,\mu)$, from
 which the lepton number susceptibilities can be extracted according 
 to \eq\nr{Xi_master}. Solid, dashed, and wiggly lines correspond to 
 fermions, scalars, and gauge fields, respectively.  
} 
\la{fig:graphs}
\end{figure}
%

A straightforward computation 
making use of \eqs\nr{tadpole_b}, \nr{tadpole_f} and
implementing the appropriate resummation
yields
\footnote{%
 Higher orders could be worked out 
 with the same techniques as employed in ref.~\cite{ag}. 
 }
\ba
 && \hspace*{-2cm} 
 \frac{12\, [ P(T,  \mu  ) - P(T,0) ]}{T^2} 
 \nn
 & = & 
 6\, \biggl[ 1 - \frac{3}{8\pi^2} \biggl(
  \frac{g_1^2}{36} + \frac{3 g_2^2}{4} + \frac{4 g_3^2}{3}
  \biggr)\biggr]
   {\rm tr } ( \mu  _ q ^ 2 ) 
   \nn
 & + & 
 3\, \biggl[ 1 - \frac{3}{8\pi^2} \biggl(
  \frac{4g_1^2}{9} + \frac{4 g_3^2}{3}
 \biggr)\biggr] 
  {\rm tr } ( \mu  _ u ^ 2 ) \nn 
 & + & 
 3\, \biggl[ 1 - \frac{3}{8\pi^2} \biggl(
  \frac{g_1^2}{9} + \frac{4 g_3^2}{3}
 \biggr)\biggr] 
  {\rm tr } ( \mu  _ d ^ 2 ) \nn 
 & + & 
 2\, \biggl[ 1 - \frac{3}{8\pi^2} \biggl(
  \frac{g_1^2}{4} + \frac{3 g_2^2}{4}
 \biggr)\biggr] 
  {\rm tr } ( \mu  _ \ell ^ 2 ) \nn 
 & + & 
 \;\; \biggl[ 1 - \frac{3}{8\pi^2} 
  g_1^2
 \biggr] 
  {\rm tr } ( \mu  _ e ^ 2 ) \nn 
 & + & 
 4\, \biggl[ 1 - \frac{3 m_\varphi}{2\pi T}
 + \frac{3}{4\pi^2} \biggl( 
  2 \lambda  + 
  \frac{g_1^2 + 3 g_2^2}{8 } 
 \biggr) 
 \biggr]\, \mu_{\varphi}^2 \nn 
 & + & 3 
 \biggl[ \frac{1}{4\pi^2}
    {\rm tr }( h^{ } _ u h _ u ^\dagger ) \mu_\varphi^2
 - \frac{3}{8\pi^2} {\rm tr} \Bigl(h _ u ^\dagger h^{ } _ u  \mu_{q}^2 
         + h^{ } _ u h _ u  ^\dagger \mu_{u}^2  \Bigr) 
 \biggr]^{ }
 \nn 
 & + & 3 
 \biggl[ \frac{1}{4\pi^2}
 {\rm tr }( h^{ } _ d h _ d ^\dagger )  \mu_\varphi^2
 - \frac{3}{8\pi^2} 
{\rm tr} \Bigl(h _ d ^\dagger h^{ } _ d  \mu_{q}^2 
         + h^{ } _ d h _ d  ^\dagger \mu_{d}^2  \Bigr) 
 \biggr]^{ }
 \nonumber \\
 & + & 
 \biggl[ \frac{1}{4\pi^2}
 {\rm tr }( h^{ } _ e h _ e ^\dagger ) \mu_\varphi^2
 - \frac{3}{8\pi^2} 
  {\rm tr} \Bigl(h _ e ^\dagger h^{ } _ e  \mu_{\ell}^2 
         + h^{ } _ e h _ e  ^\dagger \mu_{e}^2  \Bigr) 
 \biggr]^{ }
 + O(\mu^4) \;. \label{pressure}
\ea
The leading correction is the term proportional to
$ m _ \varphi /(\pi T) \sim g/\pi $. It is the leading 
contribution of
soft ($ \vec k \sim m _ \varphi  $) Higgs bosons, 
from the thermal mass resummed  
\cite{carrington} 1-loop diagram. 
Thus, whereas NLO corrections to 
the spectral function are of $ O(g ^ 2) $
in the relativistic and non-relativistic regimes, 
corrections to susceptibilities already start at $ O(g) $. 
Numerically, $m_\varphi \lsim 0.6 T$ everywhere
in the symmetric phase, so the correction is less than 30\%.
The terms of  $ O ( g ^ 2 ) $ in eq.~(\ref{pressure}) contribute
to the susceptibilities at next-to-next-to-leading order (NNLO). 
We expect that additional contributions of the same order appear at two loops
when also the gauge boson thermal masses are resummed, but  
we have not calculated these terms. 

According to \eq\nr{Xi}, the desired matrix is obtained from 
\be
 \Xi_{ab} = \frac{\partial^2 }{\partial \mu_a \partial \mu_b}
 \Bigl\{ 
 \left. P(T,\mu) \right|_{\partial P / \partial {\mu_{\bar{a}}} = 0}
 \Bigr\}
 \;. \label{Xi_master}
\ee
Relevant for us is the inverse $\Xi^{-1}$, 
cf.\ \eqs\nr{washout44.3}, \nr{pre_washout_rate}.
Let us give a few examples: 

\paragraph{(i)}

Very high temperatures
($T \gsim 10^{13}$~GeV). This case  was already 
discussed  at leading order in \se\ref{ss:susc_lo}. Here only the top
Yukawa interaction, the gauge interactions, and the Higgs self-coupling
have to be included ($h_{uij} \to h_t\, \delta_{i3}\delta_{j3}$, 
$h_{dij}\to 0$, $h_{eij}\to 0$). 
Denoting by $\mu_Y$, $ \mu  _ B $, and $ \mu  _ L $ 
the chemical potentials of
$ Y _ {  q _ 3 t \varphi  \ell _ {\! N _ 1 } } $, 
$ B _ { q _ 3 t } $, and $ L _ { N _ 1 } $,  
the chemical potentials in \eq(\ref{pressure}) are
\be
 \mu^{ }_{q_3} = \frac{\mu_Y}{6} + \frac{\mu_B}{3} \;, \quad
 \mu^{ }_{u_3} = \frac{2\mu_Y}{3} + \frac{\mu_B}{3} \;, \quad
 \mu^{ }_{\ell_1} = -\frac{\mu_Y}{2} + \mu_{L
} \;, \quad
 \mu_{\varphi} = \frac{\mu_Y}{2} \;; 
\ee
all other chemical potentials vanish. From 
\eqs\nr{pressure}, \nr{Xi_master} we obtain for
the inverse of the susceptibility
\be
 \Xi^{-1 }
 =  \frac{4}{T^2} \biggl\{ 
   1 + \frac{1}{16 \pi^2} 
   \biggl[ 
   \frac{4 (\pi T + m_\varphi )m_\varphi}{T^2} 
  +  \frac{37 g_1^2}{36}  
  + \frac{11 g_2^2}{4}  
  +  \frac{2g_3^2}{3}  
   - \, \frac{|h_t|^2}{12}  
   - 4 \lambda  
   \biggr]
 \biggr\}
 \;.  \label{i}
\ee
The leading term here agrees with that obtained from \eq\nr{X_ex0}. The 
correction represents, numerically, an increase of the washout 
rate of  about $4$\%.

\paragraph{(ii)}

As a second example, we consider the same temperature as above
($T \gsim 10^{13}$~GeV), 
but allow for three right-handed neutrinos, and consider
the evolution of three lepton densities.  
For a basis in which 
the matrix $ h $ is diagonal we have
$
 (T^{\ell}_a)^{ }_{ij} = \delta^{ }_{ai}\delta^{ }_{aj} 
$
and 
\be
 \mu_{q_3} = \frac{\mu_Y}{6} + \frac{\mu_B}{3} \;, \quad
 \mu_{u_3} = \frac{2\mu_Y}{3} + \frac{\mu_B}{3} \;, \quad
 \mu_{\ell_i} = -\frac{\mu_Y}{2} + \mu_{Li} \;,  \quad
 \mu_{\varphi} = \frac{\mu_Y}{2} \;, 
\ee
with all other chemical potentials set to zero. The same exercise
now leads to 
\ba
 \Xi^{-1}_{ } & = &
 \frac{3}{T^2} \biggl\{ 1 + \frac{3(g_1^2+3 g_2^2)}{32 \pi^2} \biggr\}
 \left( \begin{array}{ccc} 1 & 0 & 0 \\ 0 & 1 & 0 \\ 0 & 0 & 1 \end{array}
 \right) 
  \nn & + & 
  \frac{1}{T^2} \biggl\{ 
   1 + \frac{1}{ 16 \pi^2} 
   \biggl[ 
   \frac{16 (\pi T + m_\varphi )m_\varphi}{T^2} 
  \nn & & \hspace*{2.3cm}
  - \, \frac{7 g_1^2}{18} 
  - \frac{5 g_2^2}{2} 
  + \frac{8 g_3^2}{3} 
   -   \frac{ |h_t|^2}{3} 
   - 16 \lambda 
   \biggr]
 \biggr\}
 \left( \begin{array}{ccc} 1 & 1 & 1 \\ 1 & 1 & 1 \\ 1 & 1 & 1 \end{array}
 \right) 
 \;. \hspace*{5mm} \label{ii} 
\ea
The diagonal components of this matrix agree with \eq\nr{i}. 
The dissipation matrix of eq.~\nr{pre_washout_rate} becomes
$
 \gamma^{ }_{ab} = |h^{ }_{aa}|^2 \mathcal{W}(M_a)\, \Xi^{ -1}_{ab}
$. 
This is non-symmetric and non-diagonal; the non-diagonal components
determine how the lepton numbers $L_b$, $b\neq a$, influence the 
evolution of $L_a$ (cf.\ \eq\nr{eom}).

\paragraph{(iii)}

The final example is 
a ``low'' temperature ($10^2$~GeV $\lsim T \lsim 10^{5}$~GeV),  
such that all Standard Model interactions are in equilibrium.
Among them are strong sphalerons, but they have no particular
effect since the chirality
flipping processes are also mediated by the quark Yukawa interactions.
The electroweak sphalerons violate lepton
and baryon numbers.  The SM interactions conserve the  
charges $ X _ i =  L _ i - B/3 $ and  the hypercharge $ Y $. 
Unless some of the neutrino Yukawa couplings vanish,  $ H _ {\rm int } $ 
breaks all $ X _ i $-symmetries, leaving only $ Y $ conserved.
Then, 
\ba
 && 
 \mu_{q_i} = \frac{\mu_Y}{6} + \frac{\mu
}{3} \;, \quad
 \mu_{u_i} = \frac{2\mu_Y}{3} + \frac{\mu
}{3} \;, \quad
 \mu_{d_i} = -\frac{\mu_Y}{3} + \frac{\mu
}{3} \;, \nn
 &&
 \mu_{\ell_i} = -\frac{\mu_Y}{2} - \mu_{X_i} \;, \quad
 \mu_{e_i} = - \mu_Y - \mu_{X_i} \;, \quad
 \mu_{\varphi} = \frac{\mu_Y}{2} \;
 \;, 
\ea
where 
$ 
   \mu
  \equiv \fr13 \sum _ i \mu  _ {X_i} 
$, and $i = 1,2,3   $. 
The extremization in \eq\nr{Xi_master}
takes place with respect to $\mu_Y$, which then leads to 
\ba
 \Xi^{-1}_{ } & = &
 \frac{2}{T^2} \biggl\{ 1 + \frac{3(g_1^2+g_2^2)}{(4\pi)^2} \biggr\}
 \left( \begin{array}{ccc} 1 & 0 & 0 \\ 0 & 1 & 0 \\ 0 & 0 & 1 \end{array}
 \right) 
  \nn & + & 
  \frac{40 }{237 T^2} \biggl\{ 
   1 + \frac{27}{ 790 \pi^2} 
   \biggl[ 
   \frac{16 \pi m_\varphi}{T} + \frac{312 m_\varphi^2}{79 T^2}
  \nn & & \hspace*{1.3cm}
  - \, \frac{3749 g_1^2}{288} 
  - \frac{1813 g_2^2}{288} 
  +  \frac{121 g_3^2}{3} 
   -  \frac{11 |h_t|^2}{24} 
   - 16 \lambda 
   \biggr]
 \biggr\}
 \left( \begin{array}{ccc} 1 & 1 & 1 \\ 1 & 1 & 1 \\ 1 & 1 & 1 \end{array}
 \right) 
 \;. \hspace*{9mm} \label{iii}
\ea
Numerically, the correction 
appearing in the second structure of \eq\nr{iii} is 23\%, 
and it is dominated by the term proportional to $g_3^2$. 
Close to the electroweak crossover, where
$m_\varphi \sim g^2 T/\pi$, the perturbative expansion associated
with the Matsubara zero modes breaks down, 
and non-perturbative methods are needed for determining $\Xi$.

For $T \lsim 130$~GeV, the sphaleron processes violating $B+L$
are so slow that $B$ is effectively conserved~\cite{simu}.
Then $\Xi$ is a different $3\times 3$ matrix from the above. 
In a narrow temperature
range around $T\sim 130$~GeV, 
both $B$ and $L_i$ need to be treated as 
separate slow variables, and $\Xi$
is a $4\times 4$ matrix. 
For practical purposes it may be sufficient to solve separate
3-variable non-equilibrium problems in the regimes 
$T \gsim 130$~GeV and $T \lsim 130$~GeV and just match the solutions
at $T \sim 130$~GeV by requiring continuity. 

%
\section{Lepton asymmetry}
\label{sec:asymmetry} 

To get an idea on the numerical effect of radiative corrections 
we have computed the lepton asymmetry
in a scenario with $ M _ 1 = 10 ^ { 14 }$~GeV, 
and $ M _ I \gg M _ 1 $ for $ I 
\neq 1 $. This corresponds to the example
in \se\ref{ss:susc_lo} and example (i) in \se\ref{s:Xi-nlo}. 
For the  washout factor 
\begin{align} 
   K \equiv \left . \frac{ \Gamma  _ 0} { H } \right | _ { T = M _ 1 } 
   \label{K}
   , 
\end{align} 
where 
\begin{align} 
 \Gamma  _ 0 = \frac { M _ 1 } { 8 \pi  }  \sum _ i | h _ { 1i } | ^ 2 
 \label{Gamma0}
\end{align} 
is the tree-level decay rate of $ N _ 1 $, we have used $ K = 7 $.
We have
started the evolution 
with zero  initial asymmetry and thermal $ N _ 1 $-number  density
at $ T = M _ 1 $.
We have used the non-relativistic approximation~\cite{nrlg}, 
and solved  the evolution equations until $ T= M _ 1 /10 $, below which
the asymmetry hardly changes any more. 

We find that the effect  of the $ O ( g ) $ 
corrections to $ \Xi  $ on the asymmetry is about 
3\%. The order $ O ( g ^ 2 ) $ corrections 
to  $ \Xi  $ and to $ \widetilde{ \rho  } $  
are 1.3\% and 1\%, and 
the total effect of the corrections 
on the final asymmetry is $ \sim 5\% $. 

If, in contrast, a scenario like in ref.~\cite{numsm} is considered, 
in which temperatures around the weak scale play a role 
and the dynamics takes place in the ultrarelativistic regime
($M_I \ll T$), then it is clear from \fig\ref{fig:washout}
and from the discussion below 
eq.~\nr{iii} that effects of order 100\% are to be expected. 
We have not carried out numerics for this scenario, however.

%
\section{Summary and conclusions}
\label{sec:summary} 

In this paper we have obtained a relation, \eq\nr{washout44.3},  between
the lepton number washout rate relevant
for leptogenesis, and  finite temperature equilibrium correlation 
functions. The washout rate  factorizes into a real-time
spectral function, and an inverse susceptibility matrix 
which is determined by equilibrium thermodynamics 
(cf.\ \eq\nr{Xi}).
This relation does  not make use of any
particle approximation, and is valid to all orders in Standard Model 
couplings and at any temperature. 
The main approximation made is that 
we have worked to order $\hnu^2$ in neutrino Yukawa couplings, 
which should be a good approximation in many popular leptogenesis scenarios. 

We have computed explicitly the spectral function and the susceptibility
matrix to next-to-leading order (NLO)
in Standard Model couplings for  temperatures
above the electroweak crossover temperature but below 
the mass $ M _ 1 $ of 
the lightest right-handed neutrino, i.e., when the
right-handed neutrinos are non-relativistic
(150~GeV$ \, \lsim \, \pi T \ll M_1$). This is particularly relevant
for leptogenesis in the 
strong washout regime. 

We find that even in the non-relativistic regime 
there are corrections only suppressed by $O(g)$. They 
originate from Higgs effects on the susceptibility matrix, 
cf.\ \eqs\nr{i}, \nr{ii}, \nr{iii}. In contrast, 
NLO corrections to the spectral function 
are of $O(g^2)$ in this regime, cf.\ \eq\nr{c1}.
Numerically, the $O(g)$ corrections
are a few percent, except for temperatures close
to the electroweak crossover, where they can be 
substantially larger.

In the relativistic regime $M_1 \, \lsim \, \pi T$, 
the susceptibilities remain unmodified since they are insensitive
to $M_1$. In contrast, the spectral functions become increasingly
sensitive to infrared corrections, and extensive resummations
are needed for obtaining even the complete {\em leading-order} 
results for $M_1 \, \lsim \, g T$. 
We have 
shown that fortunately, the results can be inferred, after minor 
modifications, from existing computations of the 
right-handed neutrino production rate~\cite{besak,relat}. 
Numerical results are shown in \fig\ref{fig:washout}. 
The lepton number washout rate of the relativistic 
regime plays a role at the initial 
stage of the classic leptogenesis process, erasing some of the  
lepton asymmetry that is being generated when right-handed
neutrinos are produced from the Standard Model plasma, and 
would also be relevant for scenarios in which the 
right-handed neutrino masses are at or below the weak
scale~(\cite{numsm} and references therein).  

%
\section*{Acknowledgements}

We thank Nicolas Borghini and Mathias Garny for helpful discussions, 
and Marc Sangel for drawing our attention to the need for Matsubara 
zero-mode resummation in the context of susceptibilities. 
M.L was partly supported by the Swiss National Science Foundation
(SNF) under grant 200021-140234.

%
\appendix 
\renewcommand{\theequation}{\thesection.\arabic{equation}}


\end{document}